# A Synchronization Algorithm Based on Moving Average for Robust Audio Watermarking Scheme

Zhang Jin-quan and Han Bin
(*College of Information security engineering,
Chengdu University of Information Technology, Chengdu, China*)
zhjq@cuit.edu.cn，hanbin@ cuit.edu.cn

# Abstract

A synchronization code improvement scheme based on moving average is proposed for robust audio watermarking in the paper. Prior work has shown that the synchronization code scheme based on moving average is robust, but it was suitable for the same rule was adopted in embedding watermark and synchronization code, and the imperceptibility and the search efficiency isn't be paid attention to. Hence, in this paper, we improved the original scheme. The main contribution of this paper is as following: (1) improve the algorithm in surviving from desynchronization attack, (2) improve the scheme in inaudibility, (3) optimize the choice of parameters, (4) analyze the imperceptibility of the scheme, and (5) comparison of robustness and search efficiency with other synchronization code schemes. The experimental results show that the proposed watermarking scheme maintains high audio quality and is robust to common attacks such as additive white Gaussian noise, requantization, resampling, low-pass filtering, random cropping, MP3 compression, jitter attack and time scale modification. Simultaneously, the algorithm has high search efficiency and low false alarm rate.

## 1. Instruction
1.1 Reviews on synchronization code scheme

So far, many audio watermarking methods have been developed[1]. For the copyright protection of digital audio works，a scheme is usually required to resist common signal processing operations such as additive noise, resampling, MP3 compression, low-pass filtering and so on. The low-pass filter can produce different time delay due to the cut-off frequency. MP3 compression may add about 1000 samples of 0 values in the front of the audio. Different audio or different bit rate, the number of 0 points added is not exactly the same. Whether time delay or 0 value points, they will influence to find the exact location to extract the embedded watermark for some algorithms. Additionally, desynchronize operations, such as random cutting, inserting or replacing audio content, jitter attacks are also dangerous to robust watermarking algorithms. In order to ensure the accuracy of watermark extraction, synchronization measure is considered to determine the extraction range or position.

In these algorithms which can resist the desynchronization operation, various techniques were utilized, histogram shape[2], log coordinate mapping feature [3], statistics average value[4-6], logarithmic discrete cosine transform [7], low frequency coefficient of DWT [8], and others[9-11]. In these methods, the adoption of synchronization code was very widely concerned.

In those schemes which synchronization code is used, usually, a synchronization code is embedded first, then, a message followed. When the watermarked audio is attacked by desynchronized operations such as random cutting or jittering attack and so on, the exhaustive search method is often applied to find the synchronization code. In order to search the synchronization code, the maximal number of samples required to traverse is the sum of the samples to embed the synchronization code and the watermark information once. At the same time, the more the extraction algorithm is performed, the higher the false alarm rate is, for the algorithm was required to match the synchronization information from more detected bits.

In literature [10], the synchronization code is the front 12bits of the 13-bit Barker code. 16-bit/sample audio clip was used to embed message by change the low 13 bits of samples, and one bit synchronization code was embedded into one sample. From the experimental results, the robustness of the scheme is strong, and the efficiency to search synchronization code is high. For the embedding strength is great, and only one sample is justified when one bit message is embedded into, the distortion of the audio waveform is obvious, and there are "click" at the embedding position of the synchronization code, as was mentioned in literature[12] .

In literature [8], Wu etc. embedded synchronization code in the low frequency coefficient of DWT domain with QIM[13]. The synchronization code is an m-sequence with 63 bits. In the extraction phase, as long as more than 42 bits is equal, it is considered that the synchronization code is detected. Although the computational cost during searching for synchronization code can be reduced according to the method described in the paper, but because it adopted sample-by-sample searching to find the synchronization code, the efficiency is still relatively low. In the literature, Haar wavelet was applied with eight decomposition levels. So, 256 samples are involved to embedding one bit message. For the synchronization code is 63 bits, they need 256+63=319 times searching to find the synchronization code.

In literature [14], Lie etc. proposed a watermarking scheme which the embedding method of watermark is the same as the synchronization code. The algorithm exploited differential amplitude relations in each three consecutive sections of samples to represent one-bit information. In order to improve the quality of watermarked audio, the algorithm smoothed the boundary of sections, and the psychoacoustic model was applied to control the amount of watermark disturbances below the masking thresholds. In order to reduce the change of amplitude and improve the robustness of the algorithm, about 1000 samples was required to embed one-bit message. According to the literature, the extraction algorithm needed to perform about $l*(n_1+n_2)/20$ to find one synchronization mark. Here, $l$ is the number of three consecutive sections of samples, $n_1$ is the number of bits of the synchronization code, and $n_2$ is the number of bits of the watermark. In the experiments, $l$=1020 and $n_1$=20. Assume $n_2$=128, $l*(n_1+n_2)/20$=7548. The efficiency of synchronization search is relatively low.

In literature [4, 5], Wang etc. adopted the same algorithm to embed synchronization code. Their algorithm can be seen as an improvement over the Wang's approach in literature [10]. They employed Barker code in front of the watermark to locate the position where watermark was embedded. Their scheme embedded synchronization code into the statistics average value of multiple consecutive audio samples. The synchronization code algorithm has the following characteristics. Firstly, in their experiments, the number of consecutive samples is 5. So, the efficiency of searching synchronization code is high. Secondly, the maximum change of samples is 0.75 times the selected embedding strength. If the embedding rule in literature [13] is adopted, the noise will be smaller. Thirdly, they didn't present how to decide the embedding strength and the number of consecutive samples. In their experiments, for the number of consecutive samples is 5, in order to improve the robustness, the embedding strength is 0.2. So, the distortion of the waveform is relatively large, which may bring out audible noise. An improved direction of the embedding algorithm, without reducing the robustness, is to embed one-bit message on more consecutive samples and to reduce the embedding strength. But the efficiency of synchronization search will reduce, and false alarm rate will increase. Literature [15] adopted this technique, and presented the effect of embedding synchronization marks on ODG and BER.

In literature [12], the synchronization method embedded a given sequence of bits or its inverse in consecutive samples of the audio signal. It is a novel time domain synchronization technique. So, the execution of the scheme is fast enough to be used in real-time scene. In their experiments, the number of consecutive samples is 4. The number is so small that the scheme isn't robust enough for common signal processing operation. For example, number of the embedded marks in the marked signal is 117, number of the retrieved marks in the attacked signal is 26 after re-sampling 44100->22050->44100. If re-quantization (16->8->16bit), number of the retrieved marks is 33. After 10-order Butterworth low-pass filter, cut-off frequency of 10 kHz, number of the retrieved marks is only 7.

In literature [6], Authors embedded synchronization bits by adjusting the positive and negative of the mean value of multiple continuous samples. In fact, in order to avoid affecting the auditory quality of watermarked audio, the mean value is close to 0. This means that the number of consecutive samples should be sufficient. In their experiments, the number of continuous samples is 484, and the synchronization code is 16 bits, and sample-by-sample searching is used in extracting synchronization code phase, so the searching efficiency is relatively low.

We argue that, for a synchronization scheme, robustness, inaudibility and search efficiency is three important factors. In literature [10, 12], search efficiency of their synchronization scheme is high, but the noise introduced by embedding synchronization code is clear in literature [10], and the robustness is relatively poor in literature [12]. The robustness of synchronization scheme is strong in literature [6, 14], but the searching efficiency is relatively low.

An audio signal is a one-dimensional signal. So, Embedding synchronizing information in the time domain is an important aspect. At the same time, embedding information in time domain tends to have higher embedding capacity and detecting efficiency. In this paper, we improved our earlier work[16]. The proposed synchronization scheme based on moving average for robust audio watermarking achieved a compromise among robustness, inaudibility and search efficiency.

This paper is organized as follows. In Section 2, In Section 2, we introduce the concept of moving average in audio signal. Section 3 presents our synchronization code embedding and extraction strategies. Section 4 shares the improvement of our rule. In section 5, we present experimental results and performance analysis of our scheme. Finally, Section 6 concludes the paper.

1.2 Choice of synchronization code

In audio watermarking schemes, it is very wide that using the synchronization code to locate the watermark. In literature [10], the front of 12 bits of 13-bit Barker code is chosen as synchronization code. Literature [8] used the m sequence with 63 bits as synchronization code. In literature [4, 6, 12], the length of the synchronization code is 16 bits, it consists of the 13-bit Barker code concatenating 3-bit Barker code. Chaotic sequence is adopted in literature [17, 18], and the length isn't given in literature. In addition, in literature [14], the length of 20-bit synchronization code is employed, and the author did not described the type of synchronization code.

Assume the length of synchronization code is $l$ bits. If there are $t$ bits are correct, it is considered that the synchronization code is detected. Then the false alarm rate of the algorithm can be expressed as equation (1) [8]:

$$P_{FA} = \frac{1}{2^l} \sum_{k=t}^{l} C_l^k \qquad (1)$$

When exhaustive search is used to detect the synchronization code, the extracting algorithm usually employs the method that moving forward one sample, one bit message is extracted. So, the search efficiency is low, the number of extracted bits is large, and the false alarm rate is high. Therefore, for a synchronization code algorithm, higher search efficiency and lower false alarm rate is still an important except robustness.

Compared with the watermark information, the synchronization code is often very short. As described in literature[4], the style, length and the probability of "0" and "1" of synchronization code are taken into account. The length of synchronization code is especially important. The longer it is, the lower the false alarm is.

When synchronization code is detected, if $t=l$ is required in equation (1), the false alarm rate is only related to the length of the synchronization code, regardless of the type of synchronization code. That is, $P_{FA}=1/2^l$. If $t<l$ is allowed, the false alarm rate is closely related to the type of synchronization code. Since the Barker code and m sequence have very low sidelobes, they are both suitable as synchronization code.

(a) m sequence

An *m* sequence is a periodic sequence that can be generated by a linear feedback shift register. For linear feedback systems with $k$ shift registers, if the period of its output sequence is $2^k-1$, the sequence is called *m* sequence.

Assume two m-sequences $a_n$ and $b_n$ with N bits, $a_n$, $b_n \in \{-1,1\}$, $n \in [1,n]$. The cross-correlation functions of $a_n$ and $b_n$ are as follows.

$$R_{a,b}(j) = \sum_{n=1}^{N} a_n b_{n-j} \quad (2)$$

If $a_n \in \{-1,1\}$, $n \in [1,n]$ is a m sequence, $n=2^k-1$. Autocorrelation function of m sequence has the following properties:

$$R_{a,a}(j) = \begin{cases} n, & j=0 \\ -1, & 0<j<n \end{cases} \quad (3)$$

According to the formula (1), false alarm rate of the algorithm when $l=31$, $t>28$ is less than that when $l=16$, $t=16$. False alarm rate of the algorithm is about $10^{-5}$ when $l=31$, $t>28$ or $l=16$, $t=16$.

(b) Barker Code

Barker code is a binary code group. It is not a periodic sequence. The value of each symbol is +1 or -1. The autocorrelation function of an $n$-bit Barker code is as follows.

$$R_{x,x}(j) = \begin{cases} n, & j=0 \\ \pm 1, 0, & j \neq 0 \end{cases} \quad (4)$$

Only nine Barker sequences are known, all of length $n$ at most 13. It is known from the previous analysis that the shorter the synchronization code, the greater the false alarm rate is. In order to reduce the false alarm rate, in literature [4, 5], the authors concatenated the Barker code (1 1 1 1 1 -1 -1 1 1 -1 1 -1 1) and (1 1 -1) as synchronization code.

## 2. Definition of moving average in audio signal
2.1 Definition of moving average in audio signal

Moving average (MA) is a term in statistics. It is also called moving mean (MM) or rolling mean. A moving average can be viewed as an example of a low-pass filter used in signal processing.

Assume the sample number of an audio clip $X$ is $L$, and the values of them are denoted as $x_1, x_2, \ldots, x_L$. Choose an integer $b$, and the moving average $M_B$ is defined as equation (5).

$$M_{B_i} = \frac{1}{b}(x_i + x_{i+1} + \cdots + x_{i+b-1}) = \frac{1}{b} \sum_{k=i}^{i+b-1} x_k, i \in (1, L-b+1) \quad (5)$$

We gave an example to illustrate moving average in figure 1. In the example, the Sample is the audio waveform, $M_A$ and $M_B$ is the moving average sequence for a=16 and b=26 respectively.

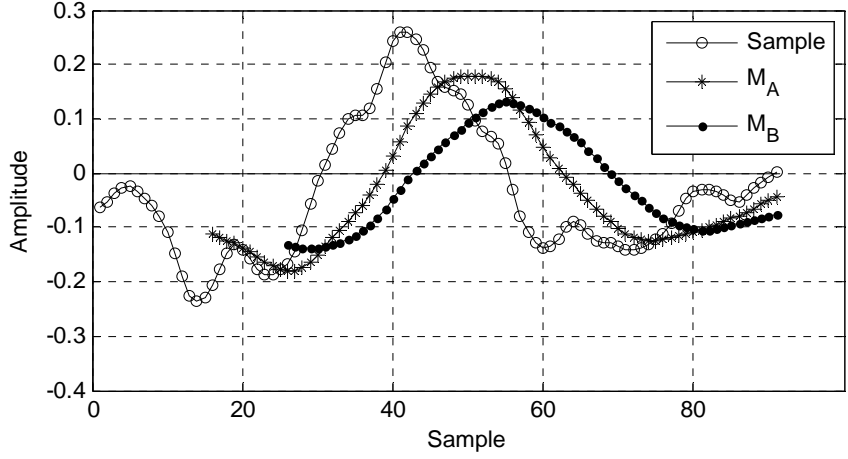

Fig. 1 Example of moving average.

2.2 Cross of two moving average

Choose two different integer *a* and *b*, *a*<*b*. then $M_A$ and $M_B$ are obtained according to equation (5). For certain $M_{B_i}$ in $M_B$, there is a corresponding $M_{A_{i+b-a}}$ in $M_A$ in time axis. If

$$(M_{B_i} - M_{A_{i+b-a}})(M_{B_{i+1}} - M_{A_{i+b-a+1}}) \leq 0, \quad i \in (1, L-b+1) \tag{6}$$

We call that there is a cross between the sequence $M_A$ and $M_B$.

2.3 Rapid calculation of moving average

The following method is used to obtain the MA of an audio segment quickly.

Assume an audio signal X=($x_1, x_2, \ldots, x_L$). For a given integer b>0,

Let

$$v_i = x_i + x_{i+1} + \cdots + x_{i+b-1}$$

Then
$$M_{B_i} = v_i/b \tag{7}$$

Now, we will compute $M_{B_{i+1}}$. In fact, compare with $M_{B_i}$, when the $M_{B_{i+1}}$ is computed, the *i*th sample is excluded, and the (*i+b*)th sample is included.

Then
$$v_{i+1} = v_i - x_i + x_{i+b}$$
$$M_{B_{i+1}} = v_{i+1}/b \tag{8}$$

The method will reduce the computation load dramatically.

In fact,
$$M_{B_{i+1}} = M_{B_i} + (x_{i+b} - x_i)/b \tag{9}$$

## 3 Synchronization code scheme

In the proposed scheme, two proper positive integers are chosen to compute the moving average sequence by sliding one sample every time. Then, the synchronization bits are embedded at crosses of the two moving average sequences with the quantization index modulation method[13]. The suggested synchronization scheme is described in the following sections.

3.1 Embedding the synchronization code

Select two different integers $a$ and $b$, $a<b$. Compute sequences $M_A$ and $M_B$ according to equation (5). The process of embedding the message is as follows.

Assume the length of the synchronization code $S$ is $l$ and the embedding strength is $s$.

Sometimes two crosses are very close. In order to improve the robustness, for the first bit of the synchronization code, the distance from position of beginning to count to the cross is greater than $b$ and without other cross. If not, the first bit of the synchronization code may not be detected. For other bits of the synchronization code, the distance from the former cross to current cross is also greater than $b$ and other crosses is allowed to be included.

The embedding process is as follows.

(1) Let $d[-1]=3s/4$, $d[1]=s/4$, $cnt=0$, $i=1$.

(2) Search a cross of $M_A$ and $M_B$ until the end of the audio clip.

do while $(M_{B_i} - M_{A_{i+b-a}})(M_{B_{i+1}} - M_{A_{i+b-a+1}}) > 0$ {

$i=i+1$,

$cnt = cnt +1$,

}

(3) If $cnt <= b$ and the first bit of the synchronization code will be embedded, then $cnt=0$, goto step (2).

(4) If $cnt < =b$, go to step (2). If not, go to step (5) to embed one bit message.

(5) Assume $u = M_{B_{i+1}}$. For the $k$th bit of synchronization code bit $S(k)$, the embedding rule is as follows:

$$u' = \begin{cases} round((u+d[1])/s) \times s - d[1], & S(k)= 1 \\ round((u+d[-1])/s) \times s - d[-1], & S(k)=-1 \end{cases} \quad (10)$$

round () denotes rounding function.

Let $d = u' - u$. Notice the $i$th point of the sequence $M_B$ corresponds to the $(i+b-1)$th sample of the audio in time axis. Those samples from former cross to the sample $x_{i+b}$, each sample adds $d$. For the first bit of the synchronization code, the starting place is the position beginning to count samples.

(6) Goto step (2) to embed the next bit.

In the embedding rule, since the amplitude of the sample of the original audio is directly modified, and the parameter *cnt* is larger than *b*, the crossed position of $M_A$ and $M_B$ will not change, as described in section 5.2.

3.2 Detecting the synchronization code

Assume the integers *a* and *b*, the embedding strength *s*, the synchronization code are obtained for the detection algorithm. Then compute sequences $M_A$ and $M_B$ according to equation (5). The detecting process of the watermark is as follows.

(1) Let *d*[-1]=3*s*/4, *d*[1]=*s*/4, *cnt*=0, *i*=1.
(2) Search a cross of $M_A$ and $M_B$ until the end of the audio clip.
do while $(M_{B_i} - M_{A_{i+b-a}})(M_{B_{i+1}} - M_{A_{i+b-a+1}}) > 0$ {
*i*=*i*+1,
*cnt* = *cnt* +1,
}
(3) If *cnt* < =*b*, go to step (2). If not, go to step (4) to extract one bit message.

(4) Assume $u' = M'_{B_{i+1}}$. The detecting rule is as follows:

$$m(k) = \begin{cases} 1, u' - \lfloor u'/s \rfloor \times s \geq s/2 \\ -1, u' - \lfloor v^*/s \rfloor \times s < s/2 \end{cases} \quad (11)$$

(5) Goto step (2) to search the next cross.

Then calculate the cross-correlation values of the synchronization code and the extracted information as follows.

$$r(k) = \sum_{i=1}^{l} S(i)m(k-l+i), k \geq l \quad (12)$$

For the synchronization code *S*, since the length of *S* is *l* bits, and the value of each bit is 1 or -1, the autocorrelation value is *l*.

Searching in *r(x)*, the position of the value *l* is the place of synchronization code.

Usually, if two positions of the value *l* are too close or too far, there is one to be ignored. The reason is that the marked audio is destroyed by cutting, adding etc., or there is a false synchronization.

From the embedding and detecting phase, we can see that, for the first bit of the synchronization code, the destination from position of beginning to count to the cross without other cross is that the first bit of the synchronization code can be detected with higher probability.

## 4. Improvement for the embedding scheme

As commented in literature [14], the proposed scheme may introduce audible noise into the watermarked audio when the embedding strength is big. literature [14] presented an idea to solve this problem, but we didn't think the idea is a good one. Fox example, it isn't proper when the value of two adjacent segment simultaneously increased or decreased. So, the core of the improvement algorithm is to make the changes of two adjacent segments as smooth as possible. In our scheme, the solution is as follows.

Let $c'$ is the change of former segment, and $c$ is that of current segment. $T_N$ samples in the front of current segment will involve in the improvement. For the first bit of synchronization code, $c' = 0$. For the sake of simplicity, we assume the variation of $T_N$ samples is linear. That is to say, it confirms with linear function $y=kx+d$, $x=0, 1, 2, …, T_N$. When $x = 0$, the sample is the last one of former segment, and the distortion of the sample is $c'$. When $x = T_N$, the distortion of the sample is $c$. So,

$$\begin{cases} c' = d \\ c = kT_N + d \end{cases} \quad (13)$$

we get the linear function is $y = (c - c')x/T_N + c'$, $x = 0,1,2,...,T_N$.

Although the improved method has little effect on the signal-to-noise ratio of marked audio, the ODG value calculated by PEAQ algorithm is improved obviously, and the robustness of the algorithm decreases very little, as described in section 5.3, 5.4 and 5.5.

## 5 Performance analysis and experiments

In experiments, we test our algorithm on different audio clips including pop music, light music, march, country music and blues music with different lengths. The experimental results are similar for all audio files tested. We report the results with three audio clips which are the pop music clip, light music clip and blues music clip. They are in WAV format, mono, 16 bits/sample, 16 s, and 44.1 kHz sampling frequency. The synchronization code is made up of 13-bit Barker code 1 1 1 1 1 -1 -1 1 1 -1 1 -1 1 concatenating 3-bit Barker code 1 1 -1.

### 5.1 Adaptive choice of parameters

The choice of parameter $b$ is related to the zero-crossing rate of the audio clip. For the efficiency of zero-crossing rate, setting a zero crossing rate threshold is a solution. Here, we adopted other method.

From the observation, the number of cross of $M_A$ and $M_B$ is close to and less than the number of zero-crossings of the audio clip. So, the value of parameter $b$ is obtained as follows.

Assume the number of zero-crossings of $M_{10}$ is $Z_{M_{10}}$, $M_{10}$ is obtained according to equation (5). $L$ is the sample number of the original audio clip. Let $num = L/Z_{M_{10}}$. The integer $b$ in embedding phase in section 3 is less than $num$ a little to resist jitter attack and time scale modification attack.

For parameter $a$, according to our experiments, for most styles of audio clips, usually, $a$ is about $2b/3$.

After the value of parameter $b$ is chosen, the embedding strength of the synchronization code can be determined in the following method.

The effect of some common signal processing operation on the audio waveform is observed in experiments.

A light music clip is chosen for low-pass filtering. A six order Butterworth filter with cut-off frequency 6kHz is used. After the low-pass filtered audio is shift left four sampling periods, its sample value subtract the sample value of the original audio correspondingly. Then, the change of each sample is obtained.

In experiments, 32 consecutive samples are chosen randomly. The waveform of original audio is show in figure 2(a), and the waveform of the low-pass filtered audio is show in figure 2(b). There are four periods time delay. Although the waveforms of figure 2(a) and figure 2(b) are almost identical, the biggest difference of them exceeds 0.01, as shown in figure 2(c). Here, the low-pass filtered audio clip shifted backward 4 periods.

To the whole audio clip, the distribution of differences of their moving average is shown in figure 2(d). When the same filter is used, the smaller the cutoff frequency of the low pass filter, the greater the noise introduced.

When additive gaussian white noise (AGWN) is used, the same experiment is made to the same audio. In experiments, the noise with 45dB SNR is chosen. As in the previous experiments, the same 32 consecutive samples are chosen. The waveform of original audio is show in figure 3(a), and the waveform of the audio with noise is show in figure 3(b). Though the waveforms of figure 3(a) and figure 3(b) are almost identical, the biggest difference of them exceeds 0.01, as shown in figure 3(c). Although the biggest difference exceeds 0.01.

To the whole audio clip, the distribution of differences of their moving average is shown in figure 3(d).

Note: figure 3(c) didn't come from figure 3(b) subtract figure 3(a), for the AWGN is performed, the figure 3(b) is different every time, even the SNR is identical. For the same reason, figure 3(d) is different every time when the AWGN is carried out and the SNR is the same.

For other signal operation, such as re-quantization, resampling, MP3 compression etc., the similar experiment can be made to obtain the differences. Embedding strength is chosen according to the distribution of differences. At the same time, if watermark and synchronization code were repeatedly embedded, the correct rate of extracting bits is only required to meet a certain value.

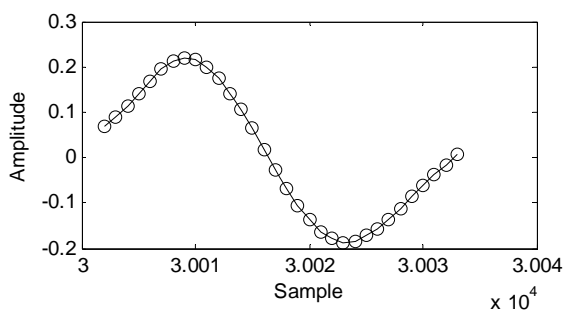

(a) Audio clip.

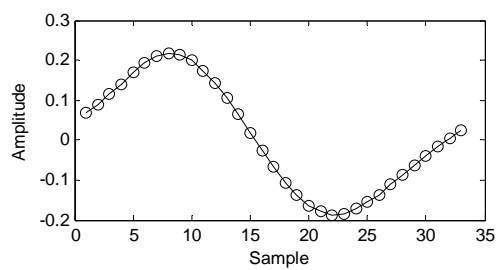

(a) Audio clip.

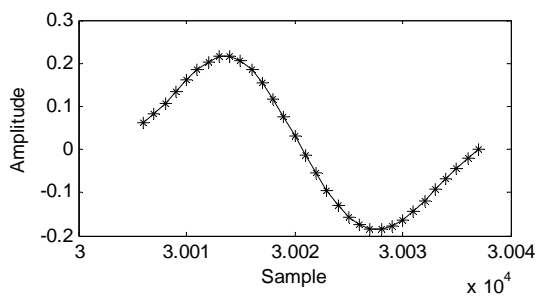

(b) Audio clip after low-pass filter.

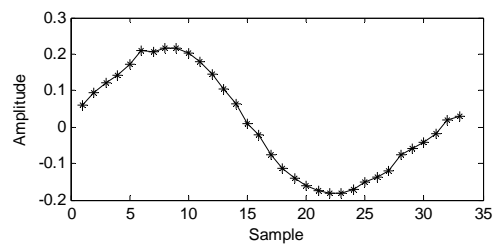

(b) Audio clip with AGWN.

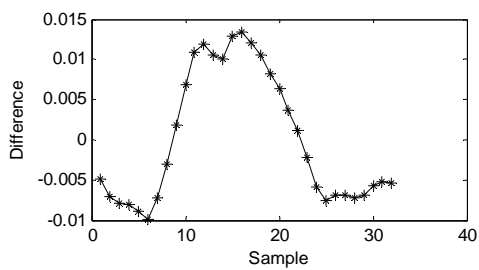

(c) Difference.

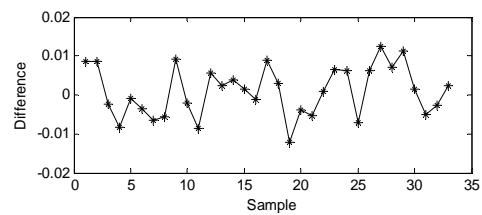

(c) Difference.

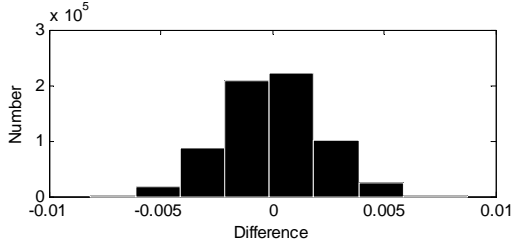
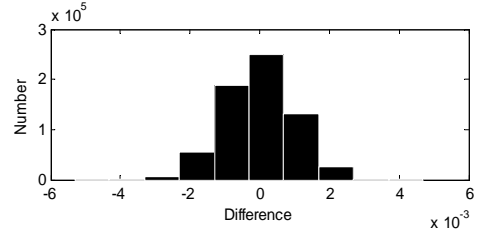

(d) Difference distribution histogram of moving average.

(d) Difference distribution histogram of moving average.

Fig. 2 Effect of the low-pass filtering.

Fig. 3 Effect of AGWN.

### 5.2 Analysis on the influence of the embedding marks to cross location

Choose two different integer $a$ and $b$, $a<b$. For certain $M_{B_i}$ in $M_B$, there is a corresponding $M_{A_{i+b-a}}$ in $M_A$. If

$$(M_{B_i} - M_{A_{i+b-a}})(M_{B_{i+1}} - M_{A_{i+b-a+1}}) \leq 0, \quad i \in (1, L-b+1)$$

There is a cross between the sequence $M_A$ and $M_B$.

Here, we assume $M_{B_i} < M_{A_{i+b-a}}$, then, $M_{B_{i+1}} > M_{A_{i+b-a+1}}$. According to equation (5), before the mark bit was embedded,

$$M_{A_{i+b-a}} = \frac{1}{a}(x_{i+b-a} + x_{i+b-a+1} + \cdots + x_{i+b-1}), \tag{14}$$

$$M_{B_i} = \frac{1}{b}(x_i + x_{i+1} + \cdots + x_{i+b-1}). \tag{15}$$

After the mark bit was embedded, according to the embedding rule,

$$M'_{A_{i+b-a}} = \frac{1}{a}((x_{i+b-a}+d) + (x_{i+b-a+1}+d) + \cdots + (x_{i+b-1}+d)) = M_{A_{i+b-a}} + d, \tag{16}$$

$$M'_{B_i} = \frac{1}{b}((x_i+d) + (x_{i+1}+d) + \cdots + (x_{i+b-1}+d)) = M_{B_i} + d. \tag{17}$$

So,

$$M'_{B_i} - M'_{A_{i+b-a}} = M_{B_i} - M_{A_{i+b-a}}. \tag{18}$$

Similarly,

$$M'_{B_{i+1}} - M'_{A_{i+b-a+1}} = M_{B_{i+1}} - M_{A_{i+b-a+1}}. \tag{19}$$

So, the crossed position of $M_A$ and $M_B$ isn't changed.

### 5.3 Analysis on SNR

In audio watermarking schemes, the SNR is a difference indicator between the watermarked and the original audio. The definition of SNR is shown as follows

$$SNR = 10\lg \frac{\sum_{i=1}^{l} x_i^2}{\sum_{i=1}^{l} (x_i - x_i^*)^2} \qquad (20)$$

where $x_i$ and $x_i^*$ are the original and marked audio signal respectively.

Assume the embedding strength is $s$. For a given audio clip, $\sum_{i=1}^{l} x_i^2$ is a constant. According to the embedding algorithm, we suppose that amplitude change of samples is a uniform distribution. That is to say, $(X^*-X) \sim U(-s/2, s/2)$, $X^*$ is the watermarked audio. So the expected value $E(X^*-X)=0$, the variance $D(X^*-X)=s^2/12$.

$$E((X^*-X)^2) = D(X^*-X) + (E(X^*-X))^2 = s^2/12. \qquad (21)$$

$$\sum_{i=1}^{L} (x_i - x_i^*)^2 \approx \sum_{i=1}^{L} s^2/12 = \frac{s^2 L}{12} \qquad (22)$$

So,

$$SNR = 10\lg \frac{\sum_{i=1}^{L} x_i^2}{\sum_{i=1}^{L} (x_i - x_i^*)^2} \approx 10\lg \frac{\sum_{i=1}^{L} x_i^2}{\frac{s^2 L}{12}} = 10\lg \frac{12 \sum_{i=1}^{L} x_i^2}{s^2 L} = 10\lg \frac{12 \sum_{i=1}^{L} x_i^2}{L} - 20\lg s \qquad (23)$$

From equation (23), The SNR is mainly affected by the embedding strength. Take the pop music for example, let $a=26$, $b=40$, the synchronization code was embedded into the audio clip repeatedly. The experimental and derivative values of the SNR are shown in Table 1 as the embedding strength changes. When other audios are used in experiments, the similar result also occurred.

$s$: the embedding strength of the synchronization code.

#ESNR: SNR from the experimental result.

#DSNR: SNR according to the derivation result.

Table 1 SNR evaluation.

| s | 0.001 | 0.002 | 0.004 | 0.006 | 0.008 | 0.01 | 0.012 | 0.014 | 0.016 | 0.018 | 0.02 |
|---|---|---|---|---|---|---|---|---|---|---|---|
| #ESNR | 54.035 | 47.933 | 41.968 | 38.439 | 35.98 | 33.962 | 32.372 | 31.155 | 29.941 | 28.952 | 28.027 |
| #DSNR | 53.949 | 47.928 | 41.908 | 38.386 | 35.887 | 33.949 | 32.365 | 31.026 | 29.867 | 28.844 | 27.928 |

SNR is an important indicator to evaluate a watermarking algorithm. Generally, the larger the embedding strength is, the larger the SNR is, the more robust the scheme is, and the more obvious the noise is. For SNR doesn't take the characteristics of human auditory into account, perceptual evaluation of audio quality (PEAQ), an assessment tool recommended by ITU BS1387, are also used in objective evaluation tests.

5.4 Comparison of the imperceptibility before and after improvement

There is an inaudibility comparison of the synchronization code method before and after improvement. In the experiments, the Barker code with 16 bits is used as synchronization code, a pseudo-random sequence with 128 bits is used as watermark message. The watermark and the synchronization code are embedded repeatedly with the same rules. In experiments, for the improved algorithm, $T_N=5$.

In comparative experiments, EA indicates that the improved algorithm isn't used for watermark and synchronization code. EB means that the improved algorithm is adopted for embedding watermark, but not for synchronization code. EC denotes the improved algorithm is employed for embedding watermark and synchronization code.

For light music, $a=16$, $b=24$, the embedding strength of the synchronization code is 0.012, and that of the watermark is 0.011. The experimental data are shown in table 2.

For pop music, $a=26$, $b=40$, the embedding strength of the synchronization code is 0.016, and that of the watermark is 0.015. The experimental data are shown in table 3.

Usually, the embedding strength of watermark is less than that of the synchronization code, for the larger the embedding strength is, the more robust the synchronization code is.

The second line(#NE): Number of embedding synchronization code in the marked signal.

The third line(#ND): Number of detecting synchronization code in the marked signal (without attacks).

The fourth line(SNR): SNR according to equation (20).

The fifth line(ODG): objective evaluation tests with PEAQ.

The fifth line(MOS): the test is done by a team composed of 10 audiences.

Table 2 Comparison of the imperceptibility before and after the improvement（light music）

|  | EA | EB | EC |
|---|---|---|---|
| #NE | 127 | 127 | 127 |
| #ND | 126 | 126 | 118 |
| *SNR*(dB) | 33.6 | 33.9 | 33.6 |
| ODG | -1.59 | -1.30 | -0.98 |
| MOS | 4.5 | 4.8 | 5.0 |

As can be seen in table 2, the synchronization code is still not completely detected without any attack. The reason is that when the synchronization code is detected, two cross correlation values with a value of 16 is close and does not meet the condition for synchronization code extraction described in section 3.2.

Table 3 Comparison of the imperceptibility before and after the improvement（pop music）

|        | EA    | EB    | EC    |
|--------|-------|-------|-------|
| #NE    | 72    | 72    | 72    |
| #ND    | 72    | 72    | 69    |
| SNR(dB)| 30.4  | 30.5  | 30.4  |
| ODG    | -2.96 | -2.25 | -0.86 |
| MOS    | 3.2   | 3.8   | 5.0   |

The reason that the marked audio has better quality after the improvement is due to the processing of the boundary. Without the improved algorithm, the difference of the samples between the marked audio and the original audio at the boundary part have a jump, as can be seen in figure 4(a). When the improved algorithm is used, the change of the samples at the boundary is relatively smooth, as can be seen in figure 4(b).

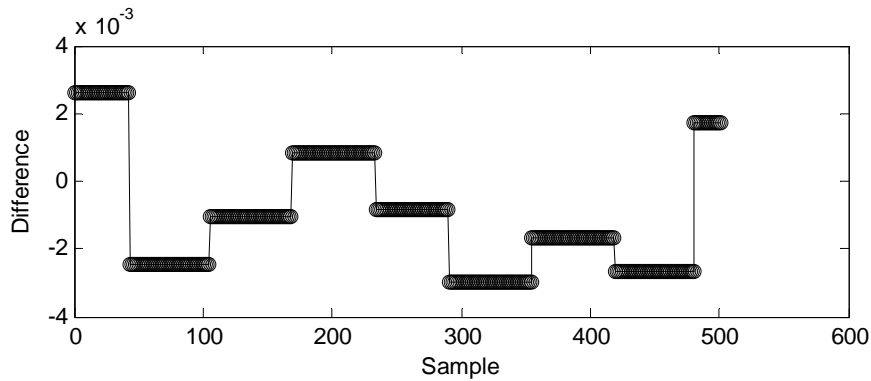

(a) Difference of the samples between the marked audio and the original audio

when the improved algorithm isn't used

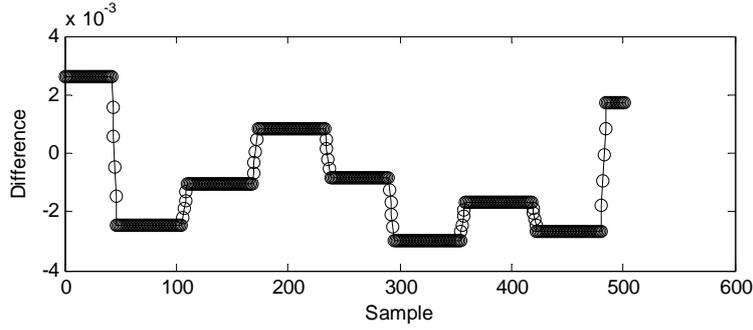

(b) Difference of the samples between the marked audio and the original audio

when the improved algorithm is used

Fig. 4 Effect of the improved algorithm to the audio samples.

Even blues music and country music, which has a lot of high frequency components, the effect of the improved algorithm is still obvious. For the blues music, $a=30$, $b=46$, the embedding strength of the synchronization code is 0.02, and that of the watermark is 0.02. The experimental data are shown in table 4.

Table 4 Comparison of the imperceptibility before and after the improvement（blues music）

|  | EA | EB | EC |
| --- | --- | --- | --- |
| #NE | 58 | 58 | 58 |
| #ND | 58 | 58 | 54 |
| SNR(dB) | 27.4 | 27.5 | 27.5 |
| ODG | -1.30 | -1.00 | -0.55 |
| MOS | 4.2 | 4.9 | 5.0 |

5.5 Robustness comparison of the scheme before and after Improvement

There is a robustness comparison of the synchronization code method for signal operation before and after the improvement. In the experiments, $T_N=5$, the value of other parameters, such as $a$, $b$ and embedding strength etc., is identical and as described in section 5.4. The experimental result is shown in table 5.

(1) Additive white Gaussian noise: white Gaussian noise is added to the marked signal.

(2) Re-quantization: the 16-bit marked audio signals are re-quantized down to 8 bits/sample and then back to 16 bits/sample.

(3) Resampling: The marked signal, originally sampled at 44.1 kHz, is re-sampled at 22.05 /11.025kHz, and then restored back by sampling again at 44.1 kHz.

(4) Low-pass filtering: A six order Butterworth filter with cut-off frequency 10kHz/8kHz/4kHz is used.

(5) Cropping: Segments of 10% are removed from the marked audio signal randomly.

(6) MP3 Compression 128kbps/96kbps/80kbps: The MPEG-1 layer-3 compression is applied. The marked audio signal is compressed at the bit rate of 128kbps/96kbps and then decompressed back to the WAVE format.

(7) Jitter attack: Randomly remove one sample out of every 1000/500/200 samples from the marked signals.

(8) time scale modification (TSM): The marked signals are scaled in time domain, where the scaling factors are $\pm 3\%, \pm 7\%, \pm 10\%$.

#ND: Number of detecting the synchronization code with the improved scheme.

#NDorg: Number of detecting the synchronization code with the original scheme.

Table 5 Robustness comparison of the scheme before and after improvement.

| Attack | Pop music | | Light music | |
|---|---|---|---|---|
| | #ND | #NDorg | #ND | #NDorg |
| No Attack | 69 | 72 | 114 | 126 |
| AWGN (55dB) | 63 | 65 | 98 | 100 |
| AWGN (45dB) | 51 | 52 | 45 | 48 |
| Requantization | 66 | 67 | 114 | 119 |
| Resampling (11025Hz) | 63 | 67 | 110 | 114 |
| Resampling (22050Hz) | 68 | 71 | 114 | 122 |
| Low-pass (10k Hz) | 58 | 55 | 41 | 42 |
| Low-pass (8k Hz) | 64 | 67 | 100 | 104 |
| Low-pass (4k Hz) | 54 | 52 | 37 | 40 |
| Cropping (10%) | 60 | 64 | 101 | 111 |
| MP3(80 kbit/s) | 38 | 31 | 10 | 11 |
| MP3(96 kbit/s) | 47 | 44 | 29 | 23 |
| MP3(128 kbit/s) | 52 | 54 | 48 | 49 |
| Jitter(1/1000) | 65 | 63 | 90 | 100 |
| Jitter(1/500) | 59 | 55 | 75 | 77 |
| Jitter(1/200) | 43 | 43 | 36 | 41 |
| TSM (-3%) | 59 | 61 | 86 | 94 |
| TSM (-7%) | 48 | 51 | 60 | 81 |
| TSM (-10%) | 45 | 44 | 56 | 57 |
| TSM (3%) | 68 | 71 | 102 | 110 |
| TSM (7%) | 64 | 67 | 77 | 84 |
| TSM (10%) | 61 | 64 | 66 | 73 |

As we can see from table 5, the robustness of the improved algorithm has a slight decrease. Compromise the robustness of the algorithm and the audibility of the marked audio, the improved algorithm is better than the original one.

5.6 Comparison of robustness with other schemes

The experimental data for light music and pop music were chosen to compare with the literature [12]. Compared with the literature [12], more samples are involved in embedding one-bit message in our scheme. As can be seen in table 6, our algorithm is more robust to the signal processing operation except MP3 compression.

#ND: the number of detecting synchronization code correctly.

# NP: the proportion of detecting synchronization code correctly.

Table 6 Comparison of robustness with other schemes

| Attack | Pop music | | Light music | | literature[12] | |
|---|---|---|---|---|---|---|
| | #ND | # NP (%) | #ND | # NP (%) | #ND | # NP (%) |
| No Attack | 69 | 96 | 118 | 93 | 117 | 100 |
| AWGN (55dB) | 63 | 88 | 99 | 78 | — | — |
| Re-quantisation | 66 | 92 | 114 | 90 | 33 | 28 |
| Re-sampling (22050Hz) | 68 | 94 | 116 | 91 | 26 | 22 |
| Re-sampling (11025Hz) | 63 | 88 | 110 | 87 | 0 | 0 |
| cropping(10%) | 60 | 83 | 104 | 82 | 105 | 90 |
| Low-pass (10kHz) | 58 | 81 | 41 | 32 | 7 | 6 |
| Low-pass (8kHz) | 64 | 89 | 102 | 80 | — | — |
| Low-pass (4kHz) | 54 | 75 | 37 | 29 | — | — |
| MP3(128kbps) | 52 | 72 | 43 | 34 | 60 | 51 |
| MP3(96kbps) | 47 | 65 | 28 | 22 | 36 | 31 |

Note:'—' means the selected schemes do not report the experimental data.

5.7 Comparison of search efficiency with other algorithms

There is a comparison of search efficiency with other algorithms. Assume the length of the synchronization code and watermark is $n_1$ bits and $n_2$ bits respectively, and the number of samples to embed the one-bit synchronization code and watermark is $l_1$ and $l_2$ respectively.

In order to find synchronization code, exhaustive search are used by all algorithms. That is, the maximum number of samples for searching synchronization code is $l_1 \times n_1 + l_2 \times n_2$.

In table 7, the expression in column 2(#ENE) is the maximum number of execution of extraction algorithm to find a synchronization code. The value of the parameters $l_1$ and $l_2$ given in the literature is shown in column 3(#VP). The number of the samples to embed one-bit synchronization code and one-bit watermark is identical in literature [6, 14] and our scheme. That is, $l_1=l_2$. For conveniently comparing, $n_1=16$ and $n_2=84$ for all algorithm. In column 4(#ND), the number of executions of the extraction algorithm is shown according to column 3.

Table 7 Comparison of search efficiency.

|      | #ENE | #VP | #ND |
|------|------|-----|-----|
| [14] | $l_1 \times (n_1+n_2)$ | $l_1=l_2=1020$ | 102,000 |
| [12] | $l_1 \times n_1 + l_2 \times n_2$ | $l_1=4, l_2=512$ | 43,072 |
| [6]  | $l_1 \times (n_1+n_2)$ | $l_1=l_2=484$ | 48,400 |
| ours | $l_2 \times n_2 / l_1 + n_1$ | $l_1=l_2<100$ | 100 |

In literature [14], the place of the synchronization code can be coarsely computed. Then, the search speed is markedly improved by almost 20 times without sacrificing the accuracy of alignment. It is still about 5000.

By comparison, we can see that our algorithm has higher search efficiency. Due to extracting less information to search for synchronization code, the false alarm rate is also lower.

5.8 Note for choosing parameter b

For the requirement of embedding the first synchronization code, the value of parameter *b* must be proper. Otherwise, some audio segments will not be embedded message into, even amplitude of these segments may be large.

Take light music for example. Embedding strength of the synchronization code and the message was described in section 5.4. The waveform of the audio clip is shown in figure 5(a). The shape in figure 5(b) and figure 5(c) is the difference between the marked audio and the original audio. For figure 5(b), the marked audio is obtained according to section 5.4, that is, a=16, b=24, and the difference shows that the message is embedded all over the audio clip. But for figure 5(c), a=26, b=36, and the difference shows that there is no message in some place.

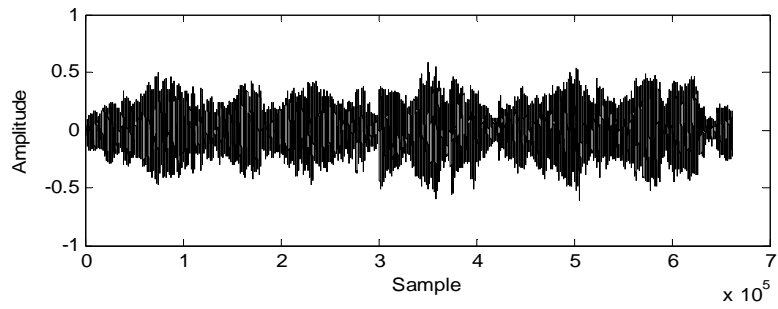

(a) waveform of the audio signal.

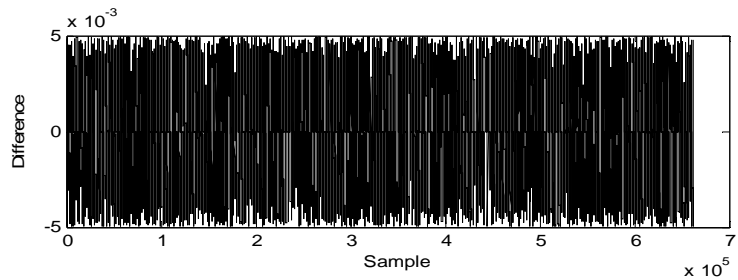

(b) Difference of marked audio and original audio when a=16, b=24.

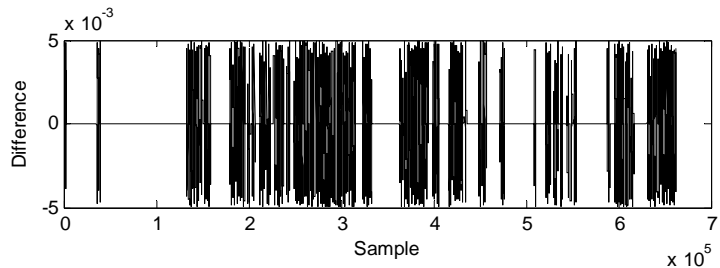

(c) Difference of marked audio and original audio when a=26, b=36.

Fig5. Effect of the choice of parameter b.

## 6. Conclusions

For moving average is robust to common signal process operation, our scheme embeds synchronization code at cross of two moving average sequences. We discuss the imperceptibility of the scheme, and compared with other algorithms in search efficiency and robustness. The experimental result shows that the proposed watermarking scheme maintains high audio quality and is robust to common attacks. Simultaneously, the algorithm has high search efficiency and low false alarm rate.

## Acknowledgments

This work is supported by the Scientific Research Foundation of CUIT under Grant No. KYTZ201420 and Scientific research project of department of education in Sichuan province under Grant No. 16ZA0221.

Thanks Prof. Hongxia Wang for valuable suggestions on improvements of this paper.